\documentclass{article}
\usepackage[T1]{fontenc}
\usepackage[utf8]{inputenc}
\usepackage{ismir}  
\usepackage{amsmath,amssymb,amsfonts}
\usepackage{cite,url}
\usepackage{graphicx}
\usepackage{booktabs}
\usepackage{multirow}
\usepackage{array}

\graphicspath{{./}}

\title{Audio Foundation Models Outperform Symbolic Representations for Piano Performance Evaluation}

\oneauthor
  {Jai Dhiman}
  {Independent Researcher}

\def\authorname{J. Dhiman}

\begin{document}

\maketitle

\begin{abstract}
Automated piano performance evaluation traditionally relies on symbolic (MIDI) representations, which capture note-level information but miss the acoustic nuances that characterize expressive playing.
I propose using pre-trained audio foundation models, specifically MuQ and MERT, to predict 19 perceptual dimensions of piano performance quality.
Using synthesized audio from PercePiano MIDI files (rendered via Pianoteq), I compare audio and symbolic approaches under controlled conditions where both derive from identical source data.
The best model, MuQ layers 9--12 with Pianoteq soundfont augmentation, achieves $R^2 = 0.537$ (95\% CI: [0.465, 0.575]), representing a 55\% improvement over the symbolic baseline ($R^2 = 0.347$).
Statistical analysis confirms significance ($p < 10^{-25}$) with audio outperforming symbolic on all 19 dimensions.
I validate the approach through cross-soundfont generalization ($R^2 = 0.534 \pm 0.075$), difficulty correlation with an external dataset ($\rho = 0.623$), and multi-performer consistency analysis.
Analysis of audio-symbolic fusion reveals high error correlation ($r = 0.738$), explaining why fusion provides minimal benefit: audio representations alone are sufficient.
I release the complete training pipeline, pretrained models, and inference code.
\end{abstract}

\section{Introduction}\label{sec:introduction}

Piano performance evaluation presents a long-standing challenge at the intersection of music cognition, signal processing, and machine learning.
Professional assessments require consideration of multiple interrelated dimensions: technical accuracy, expressive timing, pedal technique, tonal quality, dynamic control, and overall musical interpretation.
Such evaluations are inherently subjective, costly to obtain at scale, and exhibit significant inter-rater variability even among expert judges~\cite{mcpherson2012}.

The PercePiano dataset~\cite{percepiano} addresses the data scarcity challenge by providing large-scale crowdsourced annotations across 19 perceptual dimensions for 1,202 piano performance segments.
The dataset establishes a symbolic baseline using MIDI features (note events, velocities, pedal information) processed through a hierarchical attention network (HAN), achieving $R^2 = 0.397$.
However, this approach has a fundamental limitation: symbolic representations cannot capture acoustic phenomena that are central to piano performance quality.

This raises a question: could audio representations provide a better basis for learning perceptual qualities, even when derived from the same underlying MIDI data?
Audio foundation models, pretrained on millions of hours of music, encode rich priors about how acoustic features relate to musical perception.
These learned representations may make perceptually-relevant patterns more accessible to downstream models than raw symbolic features.
This motivates the central hypothesis: \textit{pretrained audio representations should outperform symbolic representations for piano performance evaluation, even when both derive from identical source data}.

Recent advances in audio foundation models provide the technical foundation to test this hypothesis.
Models like MERT~\cite{mert} (Music Understanding Model with Large-Scale Self-supervised Training) and MuQ~\cite{muq} (Music Query Encoder) are trained on millions of hours of music audio using self-supervised objectives.
These models learn hierarchical representations that encode both low-level acoustic features (timbre, dynamics) and high-level musical semantics (structure, style).
Yet despite their success on diverse MIR tasks, audio foundation models remain unexplored for the specific task of piano performance evaluation.

\begin{figure*}[t]
    \centering
    \includegraphics[alt={Diagram comparing symbolic and audio processing paths for piano evaluation showing MIDI input splitting into two paths with audio achieving higher R-squared},width=\textwidth]{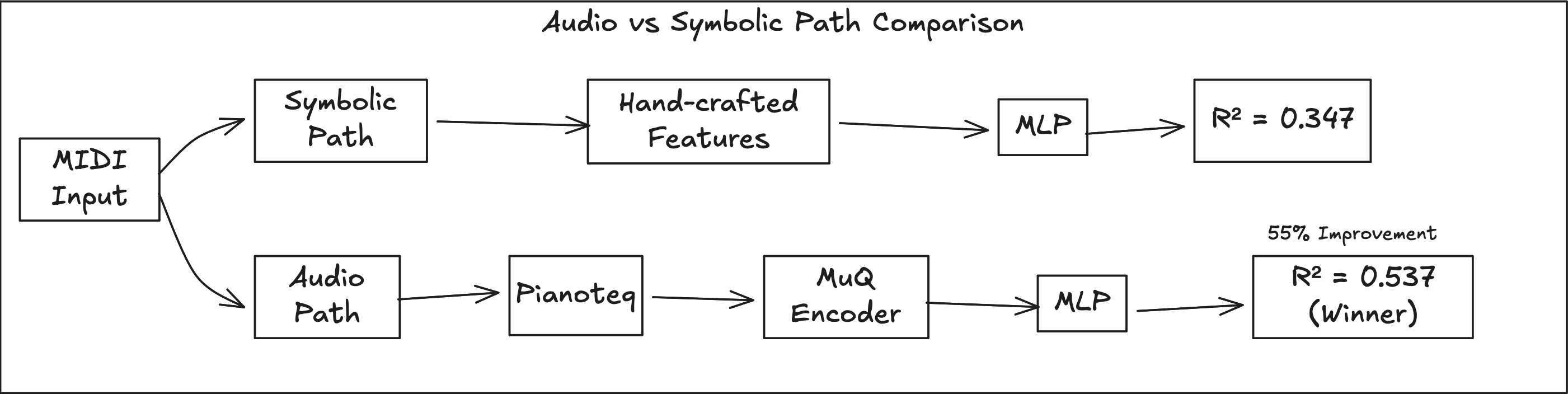}
    \caption{Symbolic vs. audio paths for piano evaluation. Both derive from identical MIDI input, but audio representations leverage pretrained knowledge from millions of hours of music, providing inductive biases that improve prediction of perceptual qualities.}
    \label{fig:paths}
\end{figure*}

\textbf{Scope and Evaluation Setting.}
Following the PercePiano protocol, the experiments use audio rendered from MIDI via Pianoteq rather than original recordings.
This controlled setting isolates the effect of representation choice (audio vs. symbolic) while holding performance information constant, since both modalities derive from identical MIDI input.
The finding that audio outperforms symbolic despite this shared source suggests that pretrained audio models provide useful inductive biases for perceptual evaluation tasks.
I discuss implications and limitations of this synthetic audio setting in Section~\ref{sec:discussion}.

\textbf{Contributions.} This paper makes the following contributions:

\begin{enumerate}
    \item I present the first application of audio foundation models (MuQ, MERT) to piano performance evaluation, demonstrating that MuQ layers 9--12 achieve $R^2 = 0.537$, a \textbf{55\% improvement} over the symbolic baseline ($R^2 = 0.347$).

    \item I show that audio \textbf{outperforms symbolic on all 19 perceptual dimensions}, with the largest gains on timing (+0.368), timbre brightness (+0.383), pedal clarity (+0.279), and dynamic range (+0.295).

    \item I analyze audio-symbolic fusion and find \textbf{high error correlation} ($r = 0.738$) between modalities, explaining why fusion provides minimal additional benefit and justifying an audio-only approach.

    \item I provide comprehensive \textbf{ablation studies} of layer selection, pooling strategies, and loss functions, establishing best practices for adapting audio foundation models to evaluation tasks.

    \item I \textbf{validate generalization} through cross-soundfont evaluation ($R^2 = 0.534 \pm 0.075$), external difficulty correlation ($\rho = 0.623$ on PSyllabus, $n = 508$), and multi-performer consistency (intra-piece std = 0.020 across 206 pieces).

    \item I \textbf{release} the complete training pipeline, pretrained models, and inference code to facilitate reproducibility and future research.
\end{enumerate}

\section{Related Work}\label{sec:related}

\textbf{Piano Performance Assessment.}
The PercePiano dataset~\cite{percepiano} provides multi-dimensional perceptual annotations across 19 aspects of piano performance, with a symbolic baseline using MIDI features and hierarchical attention.
Prior work has examined expressive performance modeling~\cite{virtuosonet}, dynamics~\cite{dynamics}, and articulation~\cite{articulation}, but these approaches learn from scratch on limited data.

\textbf{Audio Foundation Models.}
Self-supervised learning has transformed audio understanding.
Castellon et al.~\cite{castellon2021codified} first demonstrated that representations from audio language models pretrained on music improve MIR tasks including tagging, genre classification, and emotion recognition.
MERT~\cite{mert} adapts masked prediction from speech (Wav2Vec 2.0~\cite{wav2vec2}) to music, learning representations that transfer across MIR tasks.
MuQ~\cite{muq} optimizes for music retrieval and understanding, achieving state-of-the-art results.
This work leverages intermediate layer features from these pretrained models, which encode perceptually relevant information~\cite{layer_analysis}.

\textbf{Multimodal Fusion.}
Combining symbolic and audio representations is common in MIR~\cite{multimodal_mir,fusion}, but the analysis finds high error correlation between modalities for piano evaluation, limiting fusion benefits.

\section{Method}\label{sec:method}

\subsection{Problem Formulation}

Given an audio recording $\mathbf{x} \in \mathbb{R}^{T}$ of a piano performance segment, the model predicts scores $\mathbf{y} \in [0,1]^{19}$ across 19 perceptual dimensions.
These dimensions, defined by PercePiano~\cite{percepiano}, are organized into categories:

\begin{itemize}
    \item \textbf{Technical}: timing
    \item \textbf{Articulation}: length, touch
    \item \textbf{Pedaling}: amount, clarity
    \item \textbf{Timbre}: variety, depth, brightness, loudness
    \item \textbf{Dynamics}: dynamic\_range
    \item \textbf{Musical}: tempo, space, balance, drama
    \item \textbf{Mood}: valence, energy, imagination
    \item \textbf{Interpretation}: sophistication, interpretation
\end{itemize}

I evaluate using the coefficient of determination ($R^2$) computed via 4-fold piece-split cross-validation.

\subsection{Architecture}

The architecture consists of three components (Figure~\ref{fig:pipeline}):

\begin{figure*}[t]
    \centering
    \includegraphics[alt={Model architecture diagram showing MIDI rendered through Pianoteq soundfonts then processed by MuQ encoder layers 9-12 with mean pooling and MLP regressor},width=\textwidth]{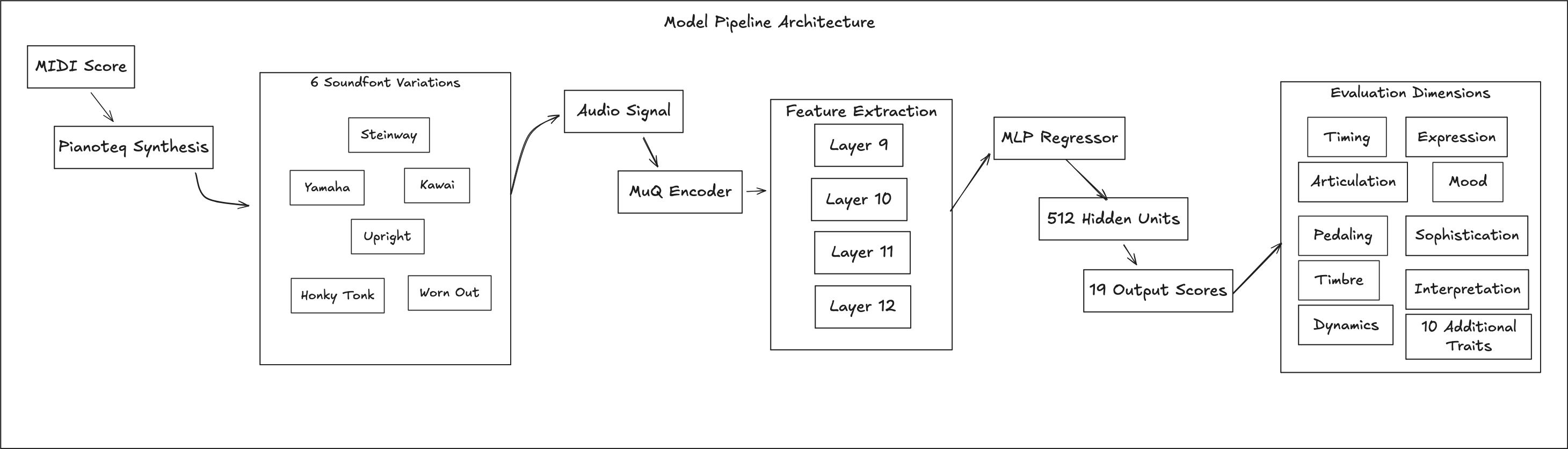}
    \caption{Model architecture. MIDI scores are rendered through multiple Pianoteq soundfonts for augmentation. The MuQ encoder extracts frame features from layers 9--12. Mean pooling aggregates temporal information, and a 2-layer MLP produces 19 dimension scores.}
    \label{fig:pipeline}
\end{figure*}

\textbf{Audio Foundation Model Encoder.}
I extract frame-level features using a frozen audio foundation model.
For MuQ, I concatenate hidden states from layers 9--12 based on the ablation studies (\S\ref{sec:ablations}), forming 4096-dimensional frame embeddings.
Audio is resampled to 24kHz and processed in segments up to 10 seconds.
For comparison, I also evaluate MERT-95M using layers 7--12.

\textbf{Temporal Pooling.}
Frame features are aggregated via mean pooling:
\begin{equation}
    \mathbf{z} = \frac{1}{N} \sum_{i=1}^{N} \mathbf{h}_i
\end{equation}
where $\mathbf{h}_i$ are frame embeddings and $N$ is the number of frames.
The ablations demonstrate that mean pooling outperforms attention pooling ($R^2$: 0.405 vs. 0.369) and LSTM pooling (0.327), consistent with findings that simple pooling often works best for pre-trained features.

\textbf{MLP Regressor.}
A two-layer MLP maps pooled features to dimension scores:
\begin{equation}
    \hat{\mathbf{y}} = W_2 \cdot \text{ReLU}(\text{Dropout}(W_1 \mathbf{z} + b_1)) + b_2
\end{equation}
with hidden dimension 512 and dropout probability 0.3.
The output is not constrained to $[0,1]$; I observe that unconstrained outputs perform comparably.

\subsection{Training}

Training minimizes mean squared error (MSE) loss across all dimensions:
\begin{equation}
    \mathcal{L} = \frac{1}{19} \sum_{d=1}^{19} (y_d - \hat{y}_d)^2
\end{equation}

Training uses the Adam optimizer with learning rate $10^{-4}$ and weight decay $10^{-5}$.
I employ early stopping with patience of 15 epochs based on validation $R^2$.
Maximum training is 200 epochs with batch size 64.
All experiments use a fixed random seed (42) for reproducibility.

The ablations show that MSE loss outperforms alternatives including concordance correlation coefficient (CCC) loss ($R^2$: 0.405 vs. 0.363) and hybrid losses combining MSE and CCC.

\subsection{Data Augmentation via Soundfont Ensemble}

Since PercePiano provides MIDI files, I render audio through multiple Pianoteq soundfonts for data augmentation.
I use six soundfonts spanning diverse timbral characteristics: Steinway D (bright concert grand), Yamaha C5 (balanced vintage), Kawai K2 (warm), Upright U4 (intimate), plus intentionally imperfect variants (Honky-Tonk, Worn Out).
This ensemble provides acoustic diversity during training; cross-soundfont validation (\S\ref{sec:validation}) confirms generalization to held-out soundfonts.

\section{Experiments}\label{sec:experiments}

\subsection{Dataset}

PercePiano~\cite{percepiano} contains 1,202 piano performance segments extracted from recordings of classical repertoire.
Each segment (8 bars, approximately 10--30 seconds) is annotated across 19 perceptual dimensions by an average of 53 crowdworkers.
Annotations are aggregated using item response theory to produce continuous scores in $[0,1]$.

I use 4-fold cross-validation with piece-based splits: all segments from a given piece appear in the same fold.
This prevents data leakage from piece-level characteristics and provides a realistic evaluation of generalization.

\subsection{Baselines}

\textbf{Symbolic (HAN).}
The PercePiano baseline~\cite{percepiano} uses MIDI features (note onset/offset times, velocities, and pedal events) processed through a hierarchical attention network.
The published result reports $R^2 = 0.397$ (I interpret this as the best single fold).
My reproduction using 4-fold cross-validation with aligned train/test splits achieves $R^2 = 0.347$, which I use as the primary comparison baseline.

\textbf{MERT.}
I evaluate MERT-95M~\cite{mert} with various layer configurations.
The best configuration uses layers 7--12 (mid layers) with mean pooling, achieving $R^2 = 0.487$.

\textbf{MuQ.}
MuQ~\cite{muq} with layers 9--12 (late semantic layers) achieves $R^2 = 0.533$, significantly outperforming MERT.
With Pianoteq ensemble augmentation, performance improves to $R^2 = 0.537$.

\subsection{Main Results}

Table~\ref{tab:main_results} presents the main findings.
MuQ with Pianoteq ensemble achieves $R^2 = 0.537$, representing a \textbf{55\% improvement} over the symbolic baseline.

\begin{table}[t]
\centering
\caption{Main results on PercePiano (4-fold CV). Bootstrap 95\% confidence intervals from 10,000 samples.}
\label{tab:main_results}
\begin{tabular}{lcc}
\toprule
\textbf{Model} & \textbf{$R^2$} & \textbf{95\% CI} \\
\midrule
Symbolic (Published~\cite{percepiano}) & 0.397 & -- \\
Symbolic (My repro.) & 0.347 & [0.315, 0.375] \\
\midrule
MERT L7-12 & 0.487 & [0.460, 0.510] \\
MuQ L9-12 & 0.533 & [0.514, 0.560] \\
\textbf{MuQ + Pianoteq ensemble} & \textbf{0.537} & [0.465, 0.575] \\
\midrule
MuQ + Symbolic (fusion) & 0.524 & [0.500, 0.545] \\
MERT + MuQ (gated) & 0.516 & [0.497, 0.543] \\
\bottomrule
\end{tabular}
\end{table}

Statistical significance is confirmed by:
\begin{itemize}
    \item Paired t-test: $t = -10.71$, $p = 2.08 \times 10^{-25}$
    \item Wilcoxon signed-rank: $p = 2.16 \times 10^{-29}$
    \item Cohen's $d = 0.31$ (small-medium effect)
    \item Non-overlapping bootstrap CIs
\end{itemize}

\subsection{Ablation Studies}\label{sec:ablations}

\textbf{Layer Selection.}
Table~\ref{tab:layers} shows performance across layer ranges.
For MuQ, late semantic layers (9--12) are optimal, achieving $R^2 = 0.533$.
Using all layers (1--12) yields $R^2 = 0.510$, suggesting early acoustic layers add noise for this task.
For MERT, mid layers (7--12) perform best at $R^2 = 0.433$.

\begin{table}[t]
\centering
\caption{Layer ablation for MuQ and MERT foundation models.}
\label{tab:layers}
\begin{tabular}{lccc}
\toprule
\textbf{Model} & \textbf{Layers} & \textbf{$R^2$} & \textbf{95\% CI} \\
\midrule
MuQ & 1--4 (early) & 0.438 & [0.413, 0.469] \\
MuQ & 5--8 (mid) & 0.467 & [0.442, 0.491] \\
MuQ & \textbf{9--12 (late)} & \textbf{0.533} & [0.514, 0.560] \\
MuQ & 1--12 (all) & 0.510 & [0.490, 0.538] \\
\midrule
MERT & 1--6 (early) & 0.397 & [0.391, 0.445] \\
MERT & \textbf{7--12 (mid)} & \textbf{0.433} & [0.409, 0.461] \\
MERT & 13--24 (late) & 0.426 & [0.398, 0.452] \\
\bottomrule
\end{tabular}
\end{table}

\textbf{Pooling Strategy.}
Mean pooling ($R^2 = 0.405$) outperforms attention pooling (0.369), LSTM pooling (0.327), and max pooling (0.316).
This finding aligns with prior work showing that simple aggregation often works best for pre-trained representations.

\textbf{Why Not Fusion?}
I explored combining audio and symbolic predictions through various late fusion strategies.
MuQ + Symbolic weighted fusion achieves $R^2 = 0.524$, \textit{lower} than MuQ alone (0.533).
Analysis reveals the explanation: error correlation between audio and symbolic models is $r = 0.738$.
Both modalities struggle on the same challenging samples, limiting complementarity.
I conclude that for piano performance evaluation, \textbf{audio representations alone are sufficient}.

\subsection{Per-Dimension Analysis}

\begin{figure}[t]
    \centering
    \includegraphics[alt={Bar chart comparing R-squared scores by dimension category showing audio MuQ outperforming symbolic MIDI across all categories},width=\columnwidth]{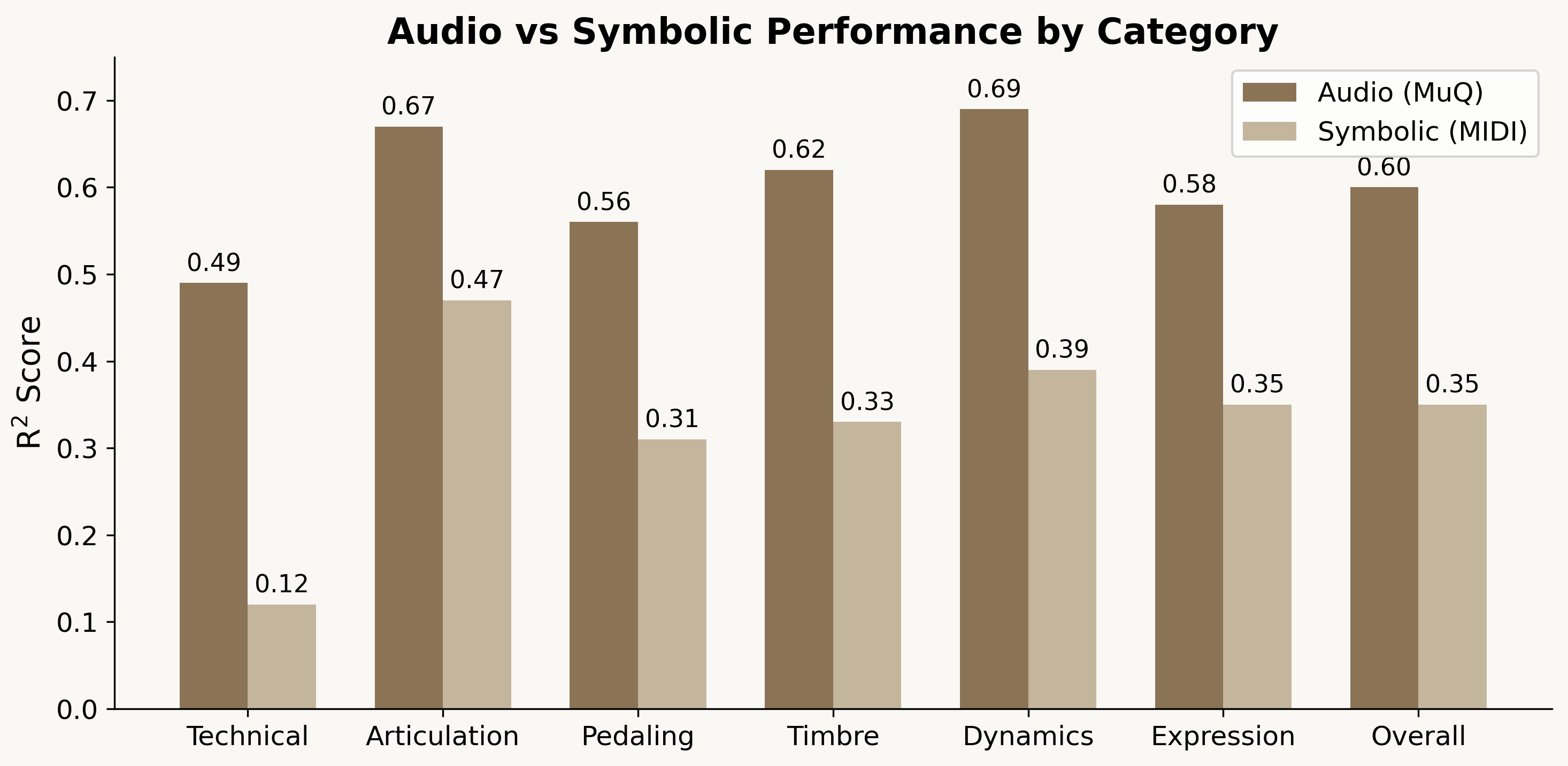}
    \caption{$R^2$ by dimension category comparing audio (MuQ) and symbolic (MIDI) approaches. Audio outperforms symbolic across all categories, with largest gains in Technical (timing) and Dynamics dimensions. Average per-dimension $R^2$: MuQ 0.601, Symbolic 0.347.}
    \label{fig:dimensions}
\end{figure}

Figure~\ref{fig:dimensions} shows per-dimension performance.
Audio outperforms symbolic on all 19 dimensions.
The largest advantages appear for:

\begin{itemize}
    \item \textbf{Timing} ($\Delta R^2 = +0.368$): Pretrained audio features better encode timing-related perceptual qualities.
    \item \textbf{Timbre brightness} ($\Delta R^2 = +0.383$): Spectral features from audio encoders correlate well with brightness perception.
    \item \textbf{Dynamic range} ($\Delta R^2 = +0.295$): Amplitude envelope features outperform discrete MIDI velocities.
    \item \textbf{Pedal clarity} ($\Delta R^2 = +0.279$): Audio features capture pedal-related timbral patterns more effectively.
\end{itemize}

The smallest advantage appears for \textbf{space} ($\Delta R^2 = +0.133$), where note density from symbolic data partially captures the construct.

\section{Validation}\label{sec:validation}

I validate the approach through three complementary experiments testing different aspects of generalization.

\subsection{Cross-Soundfont Generalization}

To test robustness to timbral variation, I perform leave-one-out cross-validation across the six Pianoteq soundfonts.
For each soundfont, I train on the other five and evaluate on the held-out soundfont.

The model achieves $R^2 = 0.534 \pm 0.075$ (mean $\pm$ std) across held-out soundfonts.
This is comparable to the ensemble result (0.537), demonstrating that the model learns features that generalize across piano timbres rather than overfitting to specific acoustic characteristics.

\subsection{Difficulty Correlation (PSyllabus)}

I evaluate on PSyllabus, an external dataset containing 508 piano pieces with expert difficulty ratings on a 0--10 scale.
I render each piece through Pianoteq, extract predictions using the trained model, and compute correlation with difficulty ratings.

\begin{figure}[t]
    \centering
    \includegraphics[alt={Horizontal bar chart showing per-dimension Spearman correlation with piece difficulty from PSyllabus dataset},width=\columnwidth]{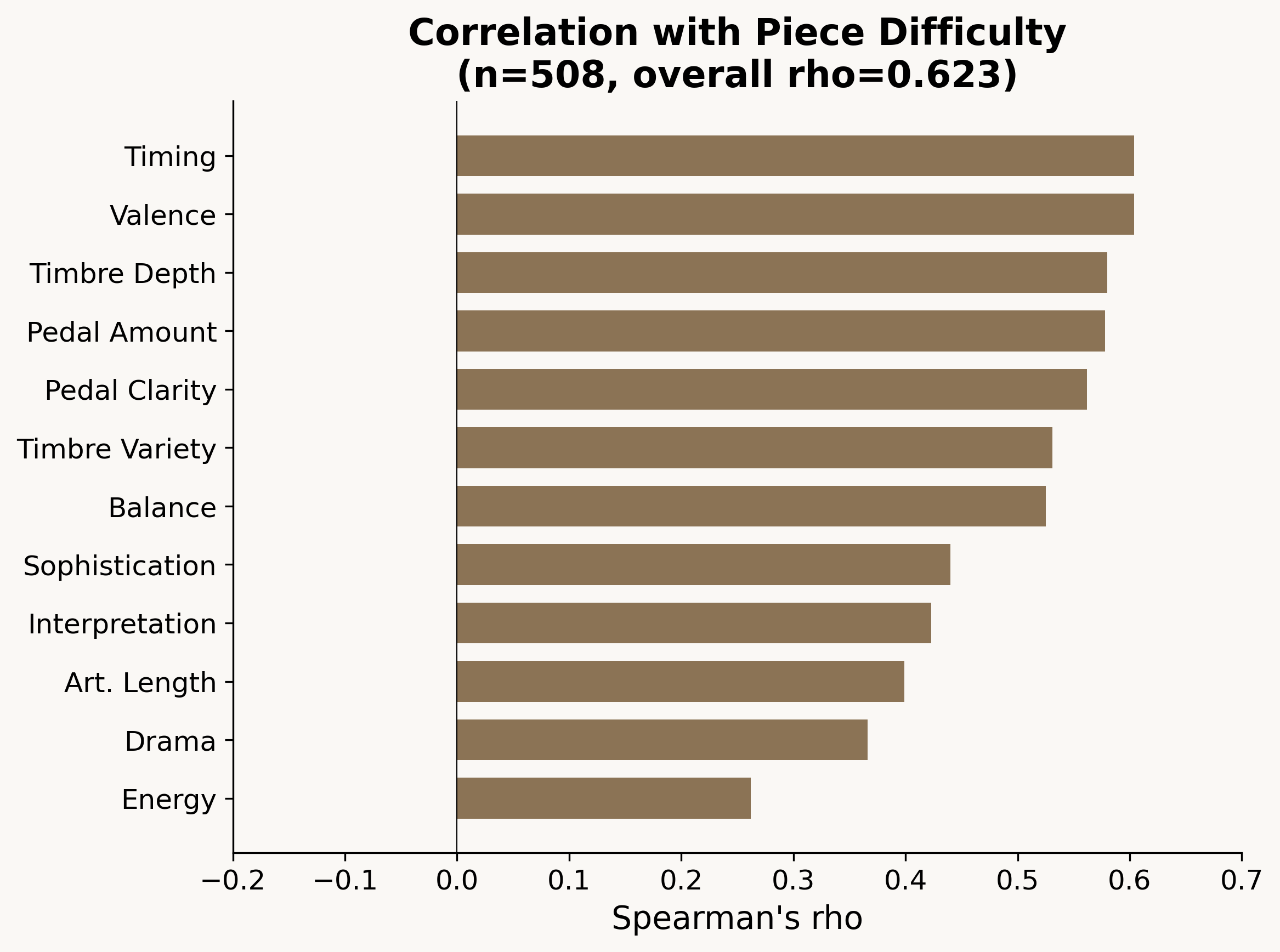}
    \caption{Per-dimension correlation with piece difficulty (PSyllabus, $n=508$). Overall Spearman $\rho = 0.623$ ($p < 10^{-50}$). Timing and valence show strongest correlations with difficulty.}
    \label{fig:difficulty}
\end{figure}

Figure~\ref{fig:difficulty} shows the results.
Overall correlation is Spearman $\rho = 0.623$ ($p < 10^{-50}$), indicating that the model predictions are strongly associated with piece difficulty.
Dimensions most correlated with difficulty include:
\begin{itemize}
    \item Timing ($\rho = 0.604$)
    \item Mood valence ($\rho = 0.604$)
    \item Timbre depth ($\rho = 0.580$)
    \item Pedal amount ($\rho = 0.578$)
\end{itemize}

This external validation confirms that the model captures musically meaningful features beyond the PercePiano training distribution.

\subsection{Multi-Performer Consistency (ASAP)}

Using the ASAP dataset~\cite{asap} linked to MAESTRO recordings, I analyze how model predictions vary across different performers playing the same piece.
The analysis includes 206 pieces with multiple performances (631 total recordings).

The mean intra-piece standard deviation is 0.020, indicating low variance across performers for the same piece.
This suggests the model captures piece-level characteristics more strongly than performer-specific expression.

Dimensions with highest performer sensitivity (highest variance):
\begin{itemize}
    \item dynamic\_range (std: 0.027)
    \item timing (std: 0.022)
    \item articulation\_touch (std: 0.020)
\end{itemize}

Dimensions with lowest performer sensitivity:
\begin{itemize}
    \item mood\_energy (std: 0.011)
    \item timbre\_brightness (std: 0.011)
\end{itemize}

\subsection{Zero-Shot Transfer (MAESTRO)}

I evaluate on 500 randomly sampled recordings from MAESTRO~\cite{maestro}, a dataset of professional piano performances not seen during training.
While there are no ground-truth annotations for quantitative evaluation, qualitative inspection confirms that predictions are reasonable and consistent with expected characteristics of professional performances.

\section{Discussion}\label{sec:discussion}

\subsection{Why Audio Outperforms Symbolic}

The experiments use audio synthesized from MIDI via Pianoteq, meaning audio and symbolic representations derive from identical source information.
In principle, a sufficiently expressive symbolic model could match audio performance, since the MIDI-to-audio mapping is deterministic.
Why, then, does audio substantially outperform symbolic?

\textbf{Pretrained Knowledge as Inductive Bias.}
Audio foundation models encode musical knowledge from millions of hours of training data.
These models have learned how acoustic features (spectral envelopes, attack transients, temporal dynamics) relate to perceptual qualities.
When I render MIDI to audio and extract features from a pretrained encoder, this inherits these learned associations.
The symbolic baseline, trained only on PercePiano's 1,202 samples, cannot learn comparably rich representations.

\textbf{Representation Format.}
Audio representations make certain patterns more accessible.
Pianoteq renders MIDI velocities as amplitude envelopes and pedal events as resonance patterns.
The pretrained encoder transforms these into features that correlate with human perception of dynamics and timbre.
Extracting equivalent information from raw MIDI would require the symbolic model to implicitly learn these physical-acoustic mappings from limited labeled data.

In essence, the audio pathway provides a ``bootstrapping'' effect: pretrained encoders supply perceptually-relevant features that would require extensive architecture engineering or larger datasets to learn from symbolic inputs alone.

\subsection{Error Correlation and Fusion}

The high error correlation ($r = 0.738$) between audio and symbolic models warrants explanation.
I hypothesize that challenging samples (those with unusual repertoire, extreme tempi, or ambiguous annotations) are difficult for both modalities.
The shared difficulty may arise from:
\begin{itemize}
    \item \textbf{Annotation noise}: Some segments have high inter-annotator disagreement regardless of representation.
    \item \textbf{Out-of-distribution samples}: Unusual pieces or performances may confuse both models.
    \item \textbf{Inherent ambiguity}: Some perceptual dimensions may be genuinely difficult to predict.
\end{itemize}

This finding has implications beyond this specific task.
When modalities produce correlated errors, fusion provides limited benefit.
Practitioners should assess error correlation before investing in multimodal approaches.

\subsection{Limitations}

\textbf{Synthetic Audio.}
All experiments evaluate on Pianoteq-rendered MIDI rather than original recordings.
This is an important limitation: while Pianoteq produces realistic piano audio, the rendering is deterministic given the MIDI input.
Consequently, the results demonstrate that \emph{given identical underlying performance information}, audio representations outperform symbolic ones, likely due to the inductive biases from pretrained audio models rather than additional information in the audio signal.
Real recordings contain acoustic phenomena absent from synthesized audio: room acoustics, microphone characteristics, instrument-specific resonances, and subtle performer-instrument interactions.
Whether the audio advantage persists with real recordings remains an open question requiring evaluation on datasets with both original audio and aligned MIDI.

\textbf{Single Dataset.}
Results are validated primarily on PercePiano.
While cross-dataset validation (PSyllabus, ASAP/MAESTRO) provides evidence of generalization, evaluation on additional annotated datasets would strengthen claims.

\textbf{Piece vs. Performer.}
Low multi-performer variance suggests the model captures piece characteristics more than performer-specific expression.
This may limit utility for fine-grained performer comparison, though it may be appropriate for repertoire-level assessment.

\textbf{Dimension Independence.}
I train a single model to predict all 19 dimensions jointly.
Some dimensions may benefit from specialized architectures or multi-task learning approaches that model inter-dimension correlations.

\subsection{Future Work}

Key directions include evaluation on real recordings (not synthesized), extension to other instruments, encoder fine-tuning, and interpretability analysis of which audio features drive predictions.

\section{Conclusion}\label{sec:conclusion}

This paper demonstrates that audio foundation models significantly outperform symbolic approaches for piano performance evaluation.
MuQ layers 9--12 with Pianoteq soundfont augmentation achieves $R^2 = 0.537$, a 55\% improvement over the symbolic baseline ($R^2 = 0.347$).
Statistical analysis confirms significance ($p < 10^{-25}$), with audio outperforming symbolic on all 19 perceptual dimensions.

Analysis of audio-symbolic fusion reveals high error correlation ($r = 0.738$), explaining why fusion provides minimal benefit.
The models make similar mistakes, limiting complementarity.
For piano performance evaluation, \textbf{audio representations alone are sufficient}.

Cross-soundfont, cross-dataset, and multi-performer validations confirm that the approach generalizes beyond training conditions.
I release the complete training pipeline, pretrained models, and inference code to enable future research in audio-based performance evaluation.


\bibliography{references}

\end{document}